\documentclass[journal]{IEEEtran}
\usepackage[normalem]{ulem}
\usepackage{float}
\usepackage{graphicx}  
\usepackage{dcolumn}      
\usepackage{amssymb}
\usepackage{quantikz}
\usepackage{amsmath}
\usepackage{amsmath}
\usepackage{multirow}
\usepackage{physics}   
\usepackage{mathtools}
\usepackage{esvect}
\usepackage{wrapfig}
\usepackage{amsthm}
\usepackage{subfig}
\usepackage{multirow}
\usepackage{makecell}
\usepackage{verbatim}
\usepackage{bbm}
\usepackage{placeins}
\usepackage[colorlinks=true,linkcolor=blue,citecolor=red,plainpages=false,pdfpagelabels]{hyperref}
\usepackage{bm}

\begin{document}

\title{KANQAS: Kolmogorov-Arnold Network for Quantum Architecture Search}

\author{
    \IEEEauthorblockN{
        Akash Kundu\IEEEauthorrefmark{1},$^{,a}$, Aritra Sarkar\IEEEauthorrefmark{2}$^{,b}$, Abhishek Sadhu\IEEEauthorrefmark{3}$^{,c}$
    }
    
    \IEEEauthorblockA{\IEEEauthorrefmark{1} QTF Centre of Excellence, Department of Physics, University of Helsinki, Finland}
    
    \IEEEauthorblockA{\IEEEauthorrefmark{1} Institute of Theoretical and Applied Informatics, Polish Academy of Sciences, Gliwice, Poland}
    
    \IEEEauthorblockA{\IEEEauthorrefmark{1} Joint Doctoral School, Silesian University of Technology, Gliwice, Poland}

\IEEEauthorblockA{\IEEEauthorrefmark{1}\IEEEauthorrefmark{2} Quantum Intelligence Research Team, Advanced Research Centre, India}

\IEEEauthorblockA{\IEEEauthorrefmark{2} QuTech, Delft University of Technology, Delft, The Netherlands}

\IEEEauthorblockA{\IEEEauthorrefmark{3} Raman Research Institute, Bengaluru, India\\}
\IEEEauthorblockA{\IEEEauthorrefmark{3} Centre for Quantum Science and Technology (CQST), International Institute of Information Technology, Hyderabad, Telangana, India}

\IEEEauthorblockA{$^a$\small\textbf{Corresponding Author}: \href{mailto:akash.kundu@helsinki.fi}{akash.kundu@helsinki.fi}}

\IEEEauthorblockA{$^b$\small: \href{mailto:a.sarkar-3@tudelft.nl}{a.sarkar-3@tudelft.nl}}

\IEEEauthorblockA{$^c$\small: \href{mailto:abhisheks@rri.res.in}{abhisheks@rri.res.in}}
}

\maketitle



\begin{abstract}
    Quantum architecture Search~(QAS) is a promising direction for optimization and automated design of quantum circuits towards quantum advantage.
    {Recent techniques in Quantum Architecture Search (QAS) emphasize Multi-Layer Perceptron (MLP)-based deep Q-networks. However, their interpretability remains challenging due to the large number of learnable parameters and the complexities involved in selecting appropriate activation functions.}
    {In this work, to overcome these challenges,} we utilize the Kolmogorov-Arnold Network (KAN) in the QAS algorithm, analyzing their efficiency in the task of quantum state preparation and quantum chemistry. 
    In quantum state preparation, our results show that in a noiseless scenario, the probability of success 
    is \textcolor{black}{$2\times$ to $5\times$} higher than MLPs. {In noisy environments, KAN outperforms MLPs in fidelity when approximating these states, showcasing its robustness against noise}. 
    In tackling quantum chemistry problems, we enhance the recently proposed QAS algorithm by integrating curriculum reinforcement learning with a KAN structure. 
    {This facilitates a more efficient design of parameterized quantum circuits by reducing the number of required 2-qubit gates and circuit depth}.
    Further investigation reveals that KAN requires a significantly smaller number of learnable parameters compared to MLPs; however, the average time of executing each episode for KAN is higher.
    
\end{abstract}

\section{Introduction}

Recent research has advanced quantum computing concepts and software development, yet significant challenges remain before real-world applications are feasible. Automating quantum algorithm design through machine learning (ML) and optimization algorithms presents promising approaches to advancing quantum hardware and programming for solving complex problems. In this context, Quantum Architecture Search (QAS)~\cite{zhang2022differentiable, lu2021markovian} inspired significantly by Neural Architecture Search~\cite{ren2021comprehensive} has been introduced.

QAS encompasses techniques to automate the optimization of quantum circuits, and it typically consists of two parts. In the first part, a template of the circuits is built where the circuit can have parameterized quantum gates, e.g., rotation angles. Then, these parameters are determined via the variational principle using a classical optimizer in a feedback loop. Algorithms constructed via this technique are called Variational Quantum Algorithms~(VQA)~\cite{mcclean2016theory,cerezo2021variational}. The parameterized circuit design in VQAs is critical, as it directly influences the expressivity and efficiency of the quantum solution. Recently, QAS has been utilized to automate the search for optimal parameterized circuits for VQAs.
QAS also finds application in determining non-parameterized circuits as an approach for quantum program synthesis~\cite{sarkar2024automated} and multi-qubit maximally entangled state preparation~\cite{kuo2021quantum}.

One of the most prominent methods to tackle QAS problems is to use Reinforcement Learning (RL) (RLQAS)~\cite{ostaszewski2021reinforcement, kuo2021quantum, kundu2024enhancing, du2022quantum, patel2024curriculum,sadhu2024qas}. In this approach, quantum circuits are defined as a sequence of actions generated by a trainable policy. The value of the cost function then optimized independently by a classical optimizer, serves as a signal for the reward function. This reward function guides the policy updates to maximize expected returns and select optimal actions for subsequent steps. Maximization of the expected return is achieved by training Neural Networks (NNs). Recently, NNs have shown potential in quantum information tasks~\cite{Morgillo2024,lumino2018experimental}. RLQAS with deep NNs can overcome the trainability issues of VQAs~\cite{WFCS+21,BK21} and has demonstrated promising outcomes in NISQ hardware~\cite{du2022quantum, patel2024curriculum}.
\begin{figure*}
	\centering
	\includegraphics[width=0.7\linewidth]{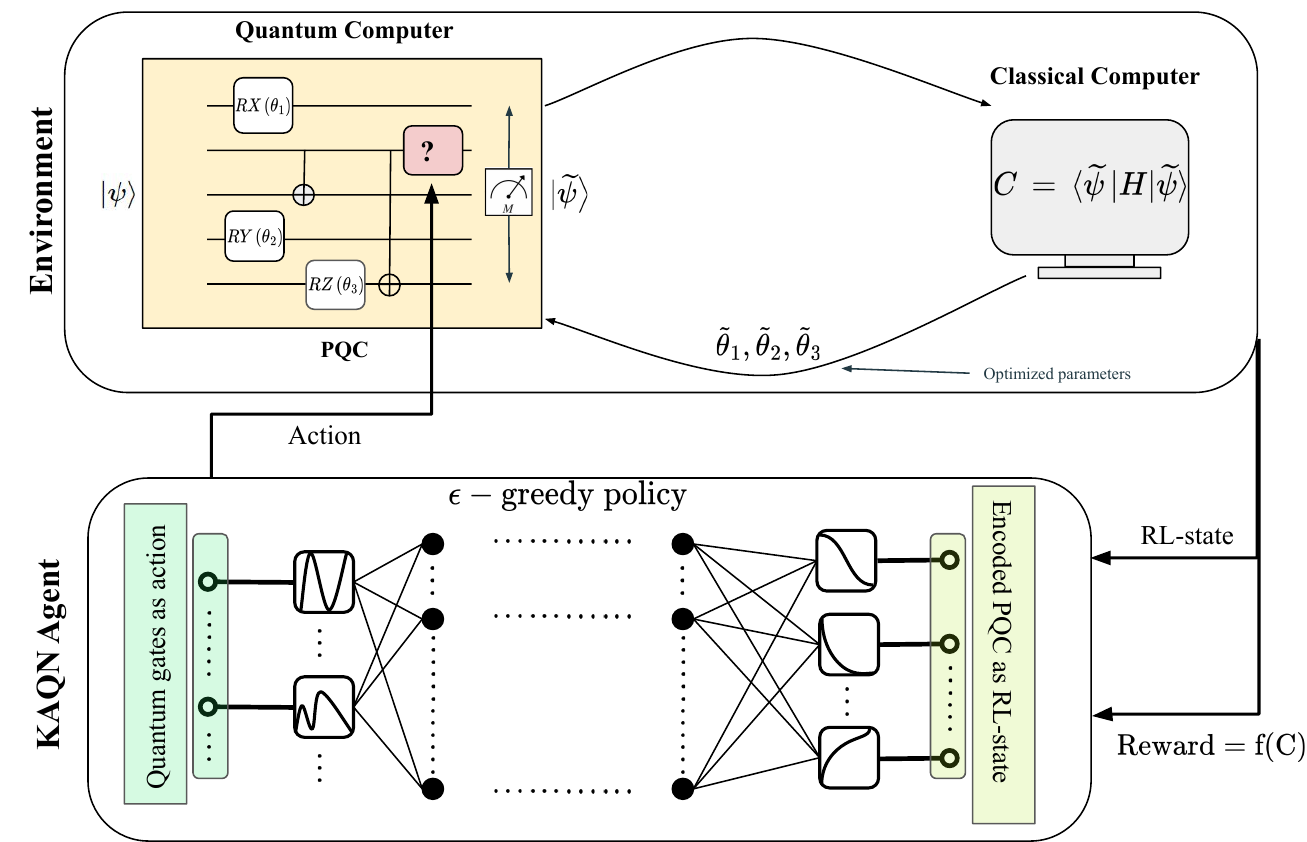}
	\caption{\small {The schematic for the KANQAS algorithm illustrates how the Kolmogorov-Arnold network replaces the traditional multi-layer perceptron in the reinforcement learning subroutine for quantum architecture search. In this setup, the environment, which incorporates a quantum algorithm, interacts with the RL-agent powered by a KAN-driven double deep-Q network, referred to as KAQN. Following an $\epsilon$-greedy policy, the agent selects its next action based on the reward function and the RL-state received from the environment. For the details of the reward function construction and quantum circuit encoding into an RL-state check the Sec.~\ref{sec:methods} and Appendix~\ref{appndix:tensor_based_encoding} respectively.}}
	\label{fig:model}
\end{figure*}

In a recent work~\cite{liu2024kan}, researchers have introduced Kolmogorov-Arnold Networks (KANs) as a novel neural network architecture typically poised to replace traditional Multi-Layer Perceptrons (MLPs). Contrary to the universal approximation theorem-based MLPs, KAN utilizes the Kolmogorov-Arnold representation theorem to approximate complicated functions. KAN has linear weights replaced by spline-based univariate functions along the network edges and is structured as learnable activation functions. Such a design enhances the accuracy and interpretability of the networks, enabling them to achieve comparable or superior results with smaller network sizes across a range of tasks, including data fitting and solving partial differential equations. Recently, different variants of KAN have been introduced~\cite{aghaei2024fkan,genet2024tkan,bozorgasl2024wav,abueidda2024deepokan,kiamari2024gkan,xu2024fourierkan}. KAN has applications in time series analysis~\cite{genet2024temporal,xu2024kolmogorov, vaca2024kolmogorov}, satellite image classification~\cite{cheon2024kolmogorov}, and improving the robustness of neural network architectures~\cite{wang2024kolmogorov}. To understand the full potential and limitations of KANs 
there is a need for further investigation towards robustness and compatibility with other deep learning architectures.

In this article, we evaluate the practicality of KANs in quantum circuit construction, analyzing their efficiency in terms of the probability of success, frequency of optimal solutions, the design of the circuit proposed by the network and their dependencies on various degrees of freedom of the network.

 To the best of our knowledge, the application of Kolmogorov-Arnold Networks in Quantum Architecture Search is still lacking in standard literature. We propose the application of KAN in QAS by \textit{replacing the MLP structure of Double Deep Q-Network (DDQN) in RLQAS with KAN to generate the desired quantum state, introducing KANQAS\footnote{\href{https://github.com/Aqasch/KANQAS_code/tree/main}{Link to KANQAS code repository}}.} As shown in Fig.~\ref{fig:model}, the proposed framework of KANQAS features an RL-agent, which contains the KAN, interacting with a quantum simulator. The agent sequentially generates output actions, which are candidates for quantum gates on the circuit. The fidelity of the state from the constructed circuit is compared to the 
quantum state fidelity of the circuit and is evaluated to determine how far it is from the desired goal. The reward, based on fidelity, is sent back to the RL-agent. This process is repeated iteratively to train the RL-agent. We show that in a noiseless scenario when constructing Bell and Greenberger–Horne–Zeilinger (GHZ) states, the probability of success and the number of optimal quantum circuit configurations to generate the states are significantly higher than MLPs. In a noisy scenario, we show that KAN can achieve a better fidelity in approximating GHZ state than MLPs, where the performance of the MLP significantly depends on the depth of the network and choice of activation function.

\textcolor{black}{Furthermore, we address quantum chemistry problems with KANQAS algorithm, where we find the ground state of 4-qubit $\texttt{H}_2$ and $\texttt{LiH}$ molecules using the recently proposed Curriculum Reinforcement Learning for Quantum Architecture Search (CRLQAS)~\cite{patel2024curriculum} as a subroutine. In our version of CRLQAS, we replace the MLP structure with KAN in the CRL. Our results indicate that KAN effectively produces a more compact parameterized quantum circuit with fewer 2-qubit gates and reduced circuit depth for identifying the ground state of the chemical Hamiltonian, while also significantly lowering the number of learnable parameters. Since 2-qubit gates typically have higher error rates than single-qubit gates~\cite{johnstun2021understanding}, this suggests that KAN could serve as a more suitable alternative to MLP for applications on real quantum hardware.}

\begin{table*}[t]
\centering

\small
\begin{tabular}{@{}ccccc@{}}
\hline
Molecule & Basis & Fermion to qubit mapping & Geometry & Number of qubits \\
\hline

$\texttt{H}_2$ & sto3g & Jordan-Wigner & {$\texttt{H}$ ($0,0,-0.35$); $\texttt{H}$ ($0,0,0.35$)    }                            & 4                \\
$\texttt{LiH}$   & sto3g & {Parity}  & {$\texttt{Li}$ ($0,0,0$); $\texttt{H}$ ($0,0,3.4$)}                                     & 4                \\
\hline
\end{tabular}
\caption{\small List of molecules included in our simulations, with configuration coordinates provided in angstroms.}
\label{tab:list_of_mol}
\end{table*}

The remainder of the paper is organized as follows. 
Sec.~\ref{sec:related_works} discussed key related developments in QAS.
We present the problem statement in Sec.~\ref{sec:problem_statement}. Meanwhile, Sec.~\ref{sec:methods} provides the methods used in this work. Specifically, we introduce the Kolmogorov-Arnold Q-network in Sec.~\ref{sec:KAQN} and discuss the RL-state, action and reward function in Sec.~\ref{sec:state_action_reward}. We present our results in Sec.~\ref{sec:results}. Where we conclude our outcome for quantum state preparation using noiseless simulations in Sec.~\ref{sec:results_noiseless}, and a noisy simulator in Sec.~\ref{sec:results_noisy}. Furthermore, in Sec.\ref{sec:results_chemistry} we present the results corresponding to finding the ground state of quantum chemistry Hamiltonian.
The resource requirements for the simulations are estimated in Sec.~\ref{sec:results_resource}. We provide concluding remarks and discuss open problems that can be tackled with KANQAS algorithm in  Sec.~\ref{sec:discussion}.

{
\section{Related works} \label{sec:related_works}
In the past few years, several key developments in the field of Quantum Architecture Search (QAS) have emerged that aim to enhance the efficiency and performance of these algorithms. The majority of these research works, in a nutshell, tackle the QAS problem by either developing sophisticated optimization strategies or utilizing well-established methods from classical machine learning such as Reinforcement Learning (RL).}

{In~\cite{zhang2021neural}, the authors introduce a framework that utilizes neural networks to predict the performance of quantum circuit architectures, significantly enhancing the efficiency of the architecture search process. In~\cite{zhang2022differentiable}, the authors utilize Monte Carlo sampling to search for Parameterized Quantum Circuits (PQCs) to solve combinatorial optimization problems.
Following this line of research, Ref.~\cite{wu2023quantumdarts} utilizes the Gumbel-Softmax sampling technique to sample quantum circuits and benchmark the QAS on quantum chemistry tasks.
In~\cite{du2022quantum}, a QAS method based on supernet and weight sharing strategy was introduced to better estimate energy for
quantum chemistry tasks. In~\cite{wang2022quantumnas}, by employing a ``SuperCircuit" encapsulating various PQCs, QuantumNAS, facilitates an efficient search for optimal PQC architecture and qubit mappings to solve quantum chemistry problems.
In \cite{he2024gradient}, the authors incorporate the information corresponding to the gradient of the PQCs, which reduces the computational burden associated with evaluating candidate architectures in QAS, allowing for more effective exploration of the design space in quantum computing tasks.
To bypass the training subroutine of different quantum architectures in~\cite{he2024training} the authors propose a proxy-based training-free QAS algorithm. Where among the introduced two proxies, the first one eliminates the unpromising PQCs and the second proxy, based on the expressibility of the PQCs, assesses their performance. 
Moreover, in~\cite{situ2024distributed}, a framework is proposed that automates the design of quantum circuit structures for interconnected quantum processing units, incorporating innovative methods for nonlocal gate implementation and qubit assignment to enhance computational efficiency.}

{Reinforcement Learning (RL) based QAS, namely RLQAS, has also been considered to automate the search for optimal PQCs. Typically, RL approaches employ a carefully designed reward function to train the agent to choose suitable gates. In~\cite{ostaszewski2021reinforcement} the authors employed Double Deep
Q-Network (DDQN) to estimate the ground state of chemical Hamiltonians. Following this line of approach in~\cite{patel2024curriculum}, the authors tackle QAS problems under realistic quantum hardware. This is achieved by introducing a tensor-based encoding for PQCs, a gate-set pruning approach, and a sophisticated second momentum-based classical optimization method, as well as by minimizing environment-agent interaction.
In~\cite{kundu2024enhancing}, by utilizing a novel encoding method for the PQCs, a dense reward function, and an $\epsilon$-greedy policy, the authors tackle the quantum state diagonalization problem. 
Additionally, in~\cite{patel2024reinforcement}, the authors show that by utilizing RL, it is possible to solve the hard instances of combinatorial optimization problems where state-of-the-art algorithms perform suboptimally. 
Finally, in~\cite{sadhu2024qas}, the authors leverage insights from quantum information theory, which helps the RL-agent to prioritize certain architectural features that are likely to provide better performance in QAS.}

{For a brief overview of QAS approaches, we encourage the authors to check the refs.~\cite{zhu2022reviewqas, martyniuk2024reviewqas, kundu2024reviewqas}.}

\section{Problem statement} \label{sec:problem_statement}

{In the previous section, we observed that significant research has focused on leveraging classical neural network architectures to improve the performance of RLQAS algorithms. While Multi-Layer Perceptron (MLP) structures in RL offer advantages, the number of trainable parameters grows rapidly with problem size, and their performance is highly sensitive to the choice of activation function. This work addresses the question: 
\textit{can Kolmogorov-Arnold Network (KAN)~\cite{liu2024kan} replace the MLP structure in RLQAS to achieve comparable or better performance while mitigating the issues of parameter growth and activation function selection?} 
We demonstrate that the answer is affirmative by tackling two crucial problems in quantum computation, which are described in the following sections.}
\textcolor{black}{\subsection{Quantum state preparation}}

To evaluate the efficiency of RL-assisted KANQAS in quantum circuit construction we check if, in a noiseless and noisy scenario, multi-qubit entanglement can be generated as expected. To this end, we target the generation of two kinds of quantum states: Bell and Greenberger–Horne–Zeilinger (GHZ) states. A Bell state is a maximal 2-qubit entangled state and is given by
\begin{equation}
    |\Phi^+\rangle = \frac{1}{\sqrt{2}} \left( |00\rangle + |11\rangle \right).
\end{equation}
The optimal circuit to produce a Bell state is given by
\begin{equation}
    \begin{quantikz}
&\gate{H}&\ctrl{1} &\qw \\
&\qw& \targ{} &\qw
\end{quantikz}
\label{fig:bell_opt_circ}.
\end{equation}
Meanwhile, a GHZ state is a 3-qubit maximally entangled state given by
\begin{equation}
    |GHZ\rangle = \frac{1}{\sqrt{2}} \left( |000\rangle + |111\rangle \right),
\end{equation}
and the optimal circuit produces a GHZ state given by
\begin{equation}
    \begin{quantikz}
&\gate{H}&\ctrl{1} & \qw&\qw \\
&\qw& \targ{} & \ctrl{1} &\qw \\
&\qw&\qw      & \targ{} &\qw.
\end{quantikz}
\label{fig:ghz_opt_circ}
\end{equation}

The objective of the KANQAS algorithm is to identify the circuit structures shown in Fig.~\ref{fig:bell_opt_circ} and Fig.~\ref{fig:ghz_opt_circ}, ideally more efficiently than an MLP-based QAS approach. As a measure of the efficiency of the network, we define the probability of success and the probability of optimal success. The probability of success is given by
\begin{equation}
\small
\frac{\text{Number of times the network finds an admissible circuit}}{\text{Total number of admissible circuits}},
\label{eq:success_equation}
\end{equation}

As the problem statement clarifies, any circuit that generates a Bell state and a GHZ state is considered an admissible circuit. Meanwhile,
Fig.~\ref{fig:bell_opt_circ} and~\ref{fig:ghz_opt_circ} are optimal admissible circuits to generate $2$- and $3$-qubit maximally entangled state. 

\textcolor{black}{\subsection{Quantum chemistry}%
Quantum chemistry~\cite{levine2009quantum} utilizes quantum mechanics to understand the behavior of atoms and molecules. Its main goal is to understand the electronic structure of chemical systems and dynamics, enabling predictions about their chemical and physical properties. This involves examining the electronic ground and excited states, reaction pathways, and transition states, crucial for explaining reactivity and interactions within chemical systems.}

\textcolor{black}{The Variational Quantum Eigensolver (VQE)~\cite{peruzzo2014variational,tilly2022variational} is a quantum-classical algorithm that is designed to find the ground state energy of quantum systems, making it particularly useful for studying molecular systems where traditional computational methods may struggle due to the complexity and the exponentially growing size of the systems.} \textcolor{black}{In VQE, the objective is to find the ground state energy of a chemical Hamiltonian $H$ by minimizing the energy 
\begin{equation}
    C(\vec{\theta}) = \min_{\Vec{\theta}}\left(\langle\psi(\Vec{\theta})|H|\psi(\Vec{\theta})\rangle \right).
    \label{eq:vqe_cost}
\end{equation}
The trial state $|\psi(\Vec{\theta})\rangle$ is prepared by applying a Parameterized Quantum Circuit (PQC), $U(\vec{\theta})$, to the initial state $|\psi_{\text{initial}}\rangle$, where $\vec{\theta}$ specify the rotation angles of the local unitary operators in the circuit.}

\textcolor{black}{The structure of the PQC is one of the crucial factors affecting the performance of the VQE. Among the recently proposed RLQAS methods for automating the search for PQC in VQE, Curriculum Reinforcement Learning (CRL) for QAS, i.e., CRLQAS~\cite{patel2024curriculum} demonstrates promising outcomes to find the ground state of various chemical Hamiltonian in the noiseless scenario and under hardware error with short PQCs.}

\textcolor{black}{Hence, to evaluate the efficiency of KAN in PQC construction for quantum chemistry problems we replace the MLP structure utilized in CRLQAS with KAN. Utilizing the CRL-assisted KANQAS we find the ground state of 4-qubit $\texttt{H}_2$ and $\texttt{LiH}$ molecules. The exact molecular structures of these molecules are provided in Table~\ref{tab:list_of_mol}.}

\section{Methods} \label{sec:methods}

\subsection{Kolmogorov-Arnold Q Networks (KAQN)} \label{sec:KAQN}



In a recent work~\cite{liu2024kan}, Kolmogorov-Arnold Networks (KAN) was proposed as a promising alternative to the Multi-Layer Perceptrons (MLP). KAN is based on the Kolmogorov-Arnold representation theorem~\cite{kolmogorov1957representation} instead of the Universal Approximation Theorem~\cite{hornik1989multilayer} used in MLP. The Kolmogorov-Arnold representation theorem states that a real-valued, smooth and continuous multivariate function $g(\textbf{t}): [0,1]^n \to \mathbbm{R}$ can be represented by a superposition of univariate functions~\cite{kolmogorov1957representation}
{
\begin{eqnarray}
    g(\mathbf{t}) = \sum_{j=1}^{2n+1} \Psi_j \bigg( \sum_{k=1}^{n} \psi_{jk} \bigg), \label{eq:KARep}
\end{eqnarray}
}
where $\Psi_j: \mathbbm{R} \to{\mathbbm{R}}$ and $\psi_{jk}: [0,1] \to{\mathbbm{R}}$. In other words, any multivariate continuous function on a bounded domain can be expressed as a composition of continuous functions of one variable. This reduces the task of learning complex multi-variable functions to learning a polynomial number of single-variable functions. It was noted by the authors of~\cite{liu2024kan} that the representation of the function in Eq.~\eqref{eq:KARep} has two layers of nonlinearity with $2n+1$ terms in the middle layer, and we need to find functions $\Psi_i$ and $\psi_{ij}$ to approximate $g(\textbf{t})$. The function $\psi_{ij}$ may be approximated using B-splines~\cite{knott99}.

A spline is a piecewise smooth polynomial function defined by a set of control points. It is defined by the order $l$ of the polynomial used to interpolate the curve between the control points. We denote the number of segments between adjacent control points as $G$. The data points are connected by the segments to form a smooth curve having $G+1$ grid points. It is observed that Eq.~\eqref{eq:KARep} can be represented as a two-layer network having activation functions at the edges and nodes performing the summation operation. However, such a network is too restrictive to approximate any arbitrary function. A way to overcome this was proposed in~\cite{liu2024kan}, where the authors propose a general architecture with wider and deeper KANs. 

The authors of~\cite{liu2024kan} define a KAN layer by a matrix $\mathbf{\Psi}$ of trainable univariate spline functions $\{\psi_{jk}(.)\}$ with $j = 1,...,n_i$ and $k = 1,...,n_o$, where $n_i$ and $n_o$ denotes the number of inputs and outputs respectively. The Kolmogorov-Arnold representation theorem can be expressed as a two-layer KAN. The inner functions constitute a KAN layer with $n_i = n$, $n_o = 2n+1$ while the outer function is another KAN with $n_i = 2n+1$, $n_o = n$. We define the shape of a KAN by $[n_1,...,n_k]$ with $k$ denoting the number of layers of the KAN. The Kolmogorov-Arnold representation theorem can be expressed as a KAN of shape $[n,2n+1,1]$. A deeper KAN can be expressed
by the composition $k$ layers:
\begin{eqnarray}
    \textbf{z} = \text{KAN}(\textbf{t}) = (\mathbf{\Psi}_k \circ \mathbf{\Psi}_{k-1} \circ ... \circ \mathbf{\Psi}_1) \mathbf{t}.
\end{eqnarray}
Since all functions are differentiable, KAN can be trained using backpropagation~\cite{RDGY13}. For the sake of simplicity, we describe a $2$-depth KAN as $[n_i, n, n_o]$, where the input layer is not included in the depth count. The output and input layers will be comprised of $n_o$, and $n_i$ nodes corresponding
to the total amount of time steps while $n$ describes the number of the hidden layers.

KAN can learn features and compositional structure due to their outer structure resembling MLPs and optimize the learned features by approximating the univariate functions with high accuracy due to their internal spline structure. It should be noted that increasing the number of layers of the dimension of the grid increases the number of parameters and, hence, the complexity of the network. 

Motivated by the developments in~\cite{KAQN2024}, we replace the MLP component of Deep Q-Networks (DQN) with the KAN. Furthermore, we employ the Double DQN (DDQN) update rule to enhance stability and learning efficiency. DDQN~\cite{van2016deep} is a Q-learning algorithm based on standard DQN~\cite{mnih2015human}, which features two neural networks to increase the stability of the prediction of Q-values for each state and action pair. For more details, see Appendix~\ref{appndx:ddqn}.

\subsection{RL-state, action space and reward function} \label{sec:state_action_reward}

\textcolor{black}{The RL environment in KANQAS is encoded using the tensor-based one-hot encoding method described in~\cite{patel2024curriculum}. However, the tensor's dimensions vary depending on the size of the action space. The RL-state encoding translates a quantum circuit into a tensor 
\begin{equation}
    \text{size} = D_\text{max}\times N\times(N+N_\text{1q})\label{eq:tensor_based_encoding_eqn},
\end{equation}
where $D_\text{max}$ is a hyperparameter and is defined as the maximum allowed gates per episode i.e. the length of an episode, $N$ is the number of qubits and $N_\text{1q}$ defines the number of 1-qubit gates. The first $N\times N$ encodes the position of the 2-qubit gate and the remaining $N\times N_\text{1q}$ encodes the position of the 1-qubit gate. {For a brief discussion on the quantum circuit encoding, see Appendix~\ref{appndix:tensor_based_encoding}}.}

{Meanwhile, the definition of the reward function varies depending on the problem under consideration. These aspects are further elaborated in the following sections.}
\vspace{1pt}
\paragraph{\textbf{\textcolor{black}{Quantum state preparation}}} 
In this problem, we initialize with an empty quantum circuit (i.e., without gates). Depending on a fidelity-based reward function of the form
\begin{equation}
        R =  
    \begin{cases}
        \mathcal{R},& \text{if } F(s_t)\geq 0.98\\
        F(s_t),              & \text{otherwise}
    \end{cases}
    \label{eq:QSP_reward_function}
\end{equation}
and by following an $\epsilon$-greedy policy the RL-agent sets the probability of selecting a random action. Where $F(s_t)$ is the fidelity of a state at step $t$ generated by KANQAS and $\mathcal{R}>>F(s_t)$, is a hyperparameter.

The random action is chosen from a predefined action space $(\mathbb{A})$ which contains non-parametrized $1$- and $2$-qubit gates~\cite{kuo2021quantum} 
\begin{equation}
    \mathbb{A} = \{CX, X, Y, Z, H, T\}.
\end{equation}
Depending on the action the RL-state, which is encoded into a tensor of dimension $D_\text{max}\times N\times(N+5)$, is modified in the next step. {For further elaboration, check the detailed discussion of the encoding of PQC provided in Appendix~\ref{appndix:tensor_based_encoding}}.
\begin{equation}
\small
R= \begin{cases}5 & \text { if } C_t<\xi, \\ 
-5 & \text { if } t \geq D_\text{max} \text { and } C_t \geq \xi, \\ 
\max \left(\frac{C_{t-1} - C_t}{C_{t-1} - C_{\min }},-1\right) & \text { otherwise }\end{cases}
\end{equation}
\textcolor{black}{\paragraph{\textbf{Quantum chemistry}}%
Following the same principle as discussed in the previous section, in this problem, we initialize with an empty quantum circuit (i.e. without any gates). However, for a fair comparison with MLP bases CRLQAS, we utilize a reward function of the form where $C_t$ is the VQE cost function (see Eq.~\ref{eq:vqe_cost}) at RL step $t$ and $\xi$ is a hyperparameter, but for VQE it is the chemical accuracy $0.0016$ Hartree. {$D_\text{max}$ is a hyperparameter that defines the maximum number of actions per episode, i.e. length of an RL-episode}.
The random action is chosen from a predefined action space $(\mathbb{A})$ which contains parametrized $1$- and non-parameterized $2$-qubit gates~\cite{patel2024curriculum, kundu2024enhancing, ostaszewski2021reinforcement} i.e. $\mathbb{A} = \{CX, RX, RY, RZ\}$.
Depending on the action, the RL-state, which is encoded into a tensor of dimension $D_\text{max}\times N\times(N+3)$, is modified in the next step. {An elaborated discussion of the encoding is provided in Appendix~\ref{appndix:tensor_based_encoding}.}}
\begin{figure}[H]
    \centering
    \subfloat[\centering ]{{\includegraphics[width=0.4\textwidth]{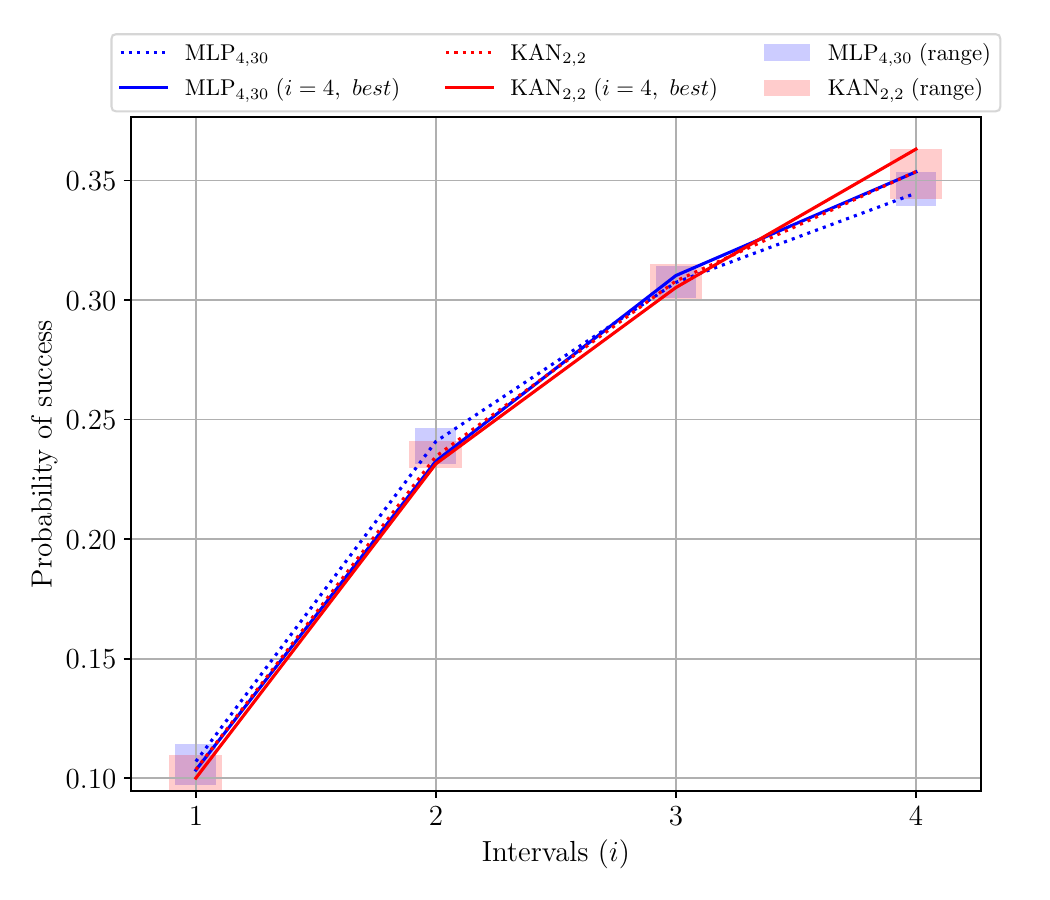} }}%
    \vspace{1pt}
    \subfloat[\centering ]{{\includegraphics[width=0.4\textwidth]{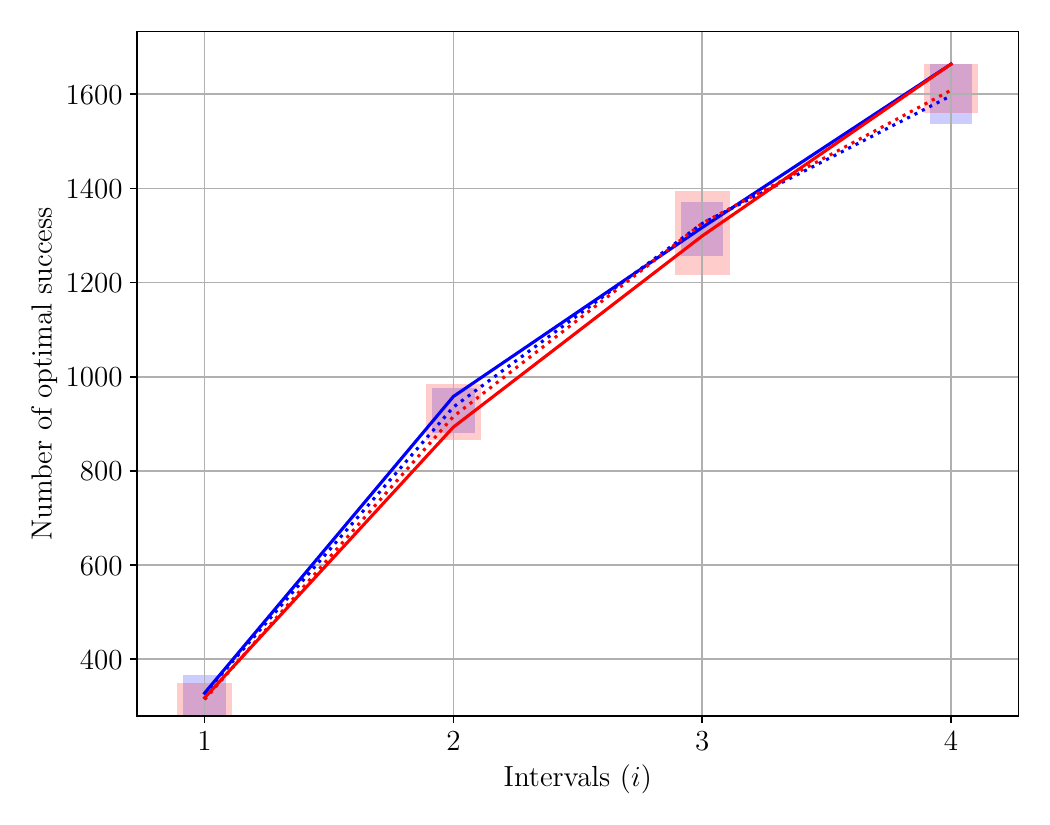} }}%
    \caption{\small \textcolor{black}{In (a) the probability of successful circuits and in (b) the probability of optimal successful circuits in finding a 2-qubit maximally entangled state is slightly higher with KAN than MLP. A total of 10000 episodes are divided into 4 separate intervals where each interval contains 2500 episodes. In (a), each point in an interval corresponds to the probability of occurrence of a successful episode (see Eq.~\ref{eq:success_equation}). Similarly, (b) corresponds to the number of occurrences of an optimal successful episode. The results are averaged over 20 random seeds (i.e. initialization) of the networks.}}%
    \label{fig:bell_state_prep_res}%
\end{figure}

\paragraph{KAN and MLP configuration notation}
{
Throughout the paper, the configurations of KANs and MLPs are written using the short notation: {$\text{KAN}_{l+1,n_h}$ for a KAN and $\text{MLP}_{l+1,n_h}$ for an MLP where $l$ number of hidden layers in the neural network, and $n_h$ is the number of neurons per hidden layer. For example, a KAN with the structure $[n_i,n^{(1)}_h,n^{(2)}_h,\ldots,n^{(l)}_h,n_o]$ is written as $\text{KAN}_{l+1,n_h}$, where the depth of the network is $l+1$ and each hidden layer has $n_h$ neurons. $n_i$ is calculated using the Eq.~\ref{eq:tensor_based_encoding_eqn}, for an example while tackling the 2-qubit ($N=2$) state reconstruction problem, we consider an action space consisting of five 1-qubit gates i.e. $N_\text{1q}=5$ and one 2-qubit gate with the maximum achievable depth $D_\text{max}=6$. Hence, following Eq.~\ref{eq:tensor_based_encoding_eqn}, we get $n_i=84$. To solve the problem, we consider a KAN of depth $2$, and the number of neurons per hidden layer is $n_h = 3$, i.e. $[84,3,12]$. The size of the output is 12 because, for 2-qubit, the \texttt{CX} gate has two variants $\texttt{CX}_{12}$ and $\texttt{CX}_{21}$ where the first index defines the control and second index is the target qubit. Meanwhile, as the 1-qubit gates can be added on both qubits so, for five 1-qubit gates, we have $10$ actions, hence a total of $12$ actions.}}

\section{Results} \label{sec:results}
{In this section, we benchmark the performance of KANQAS for quantum state preparation and quantum chemistry tasks and show that, in many instances, a KAN configuration outperforms an MLP with fewer trainable parameters. For the quantum state preparation, we consider the task of finding a maximally entangled state of $2$- and $3$-qubit in a noiseless and noisy scenario. Meanwhile, we use the KAN and MLP structures for quantum chemistry to find the ground state of $4$-qubit $\texttt{H}_2$ and $\texttt{LiH}$ molecules.}

\subsection{Quantum state preparation}

\subsubsection{Noiseless simulation} \label{sec:results_noiseless}
{In the noiseless case, we consider the structure of the KAN and the MLP as presented in Table~\ref{tab:noiseless_state_prep_network_config}. One of the prominent advantages of KAN is that its activation function is learnable; meanwhile, in the case of MLP, the activation function is a hyperparameter. In the upcoming sections, we will see that the performance of the MLP is heavily dependent on the choice of the activation function.} 

{For the construction of Bell state, we run a total of $10000$ episodes where, within each episode, we allow the agent to create a quantum circuit of maximum depth $6$, i.e. $D_\text{max}=6$. For a better representation of the results, we divide the total number of episodes into 4 intervals, where each interval contains $2500$ episodes. Fig.~\ref{fig:bell_state_prep_res}(a) investigates the probability of success in each interval, which is calculated using the Eq.~\ref{eq:success_equation}. Meanwhile, in Fig.~\ref{fig:bell_state_prep_res}(b), we see the variation in the number of optimal admissible circuits in each interval. We call a quantum circuit an admissible circuit when the state generated by it obtains a positive reward based on the reward function in Eq.~\ref{eq:QSP_reward_function}.}
\begin{figure}[H]
	\centering
	\subfloat[\centering ]{{\includegraphics[width=0.4\textwidth]{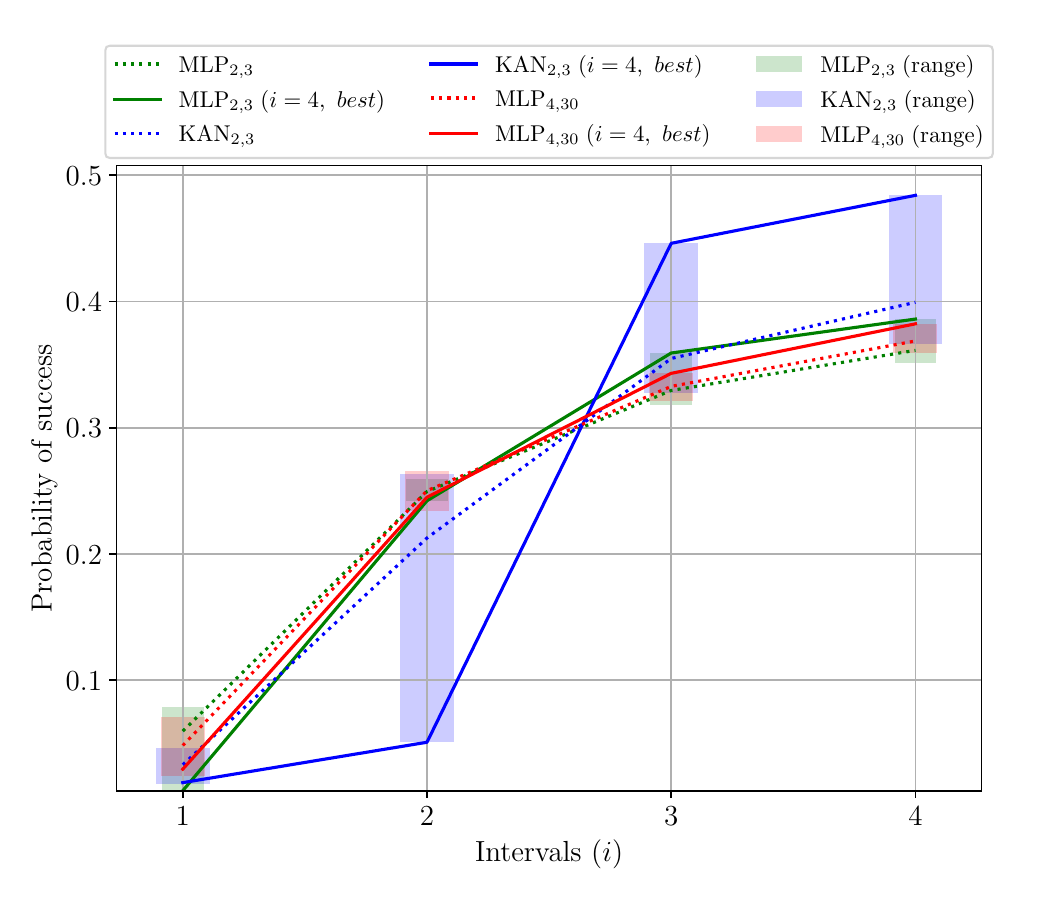} }}%
	\qquad
	\subfloat[\centering ]{{\includegraphics[width=0.4\textwidth]{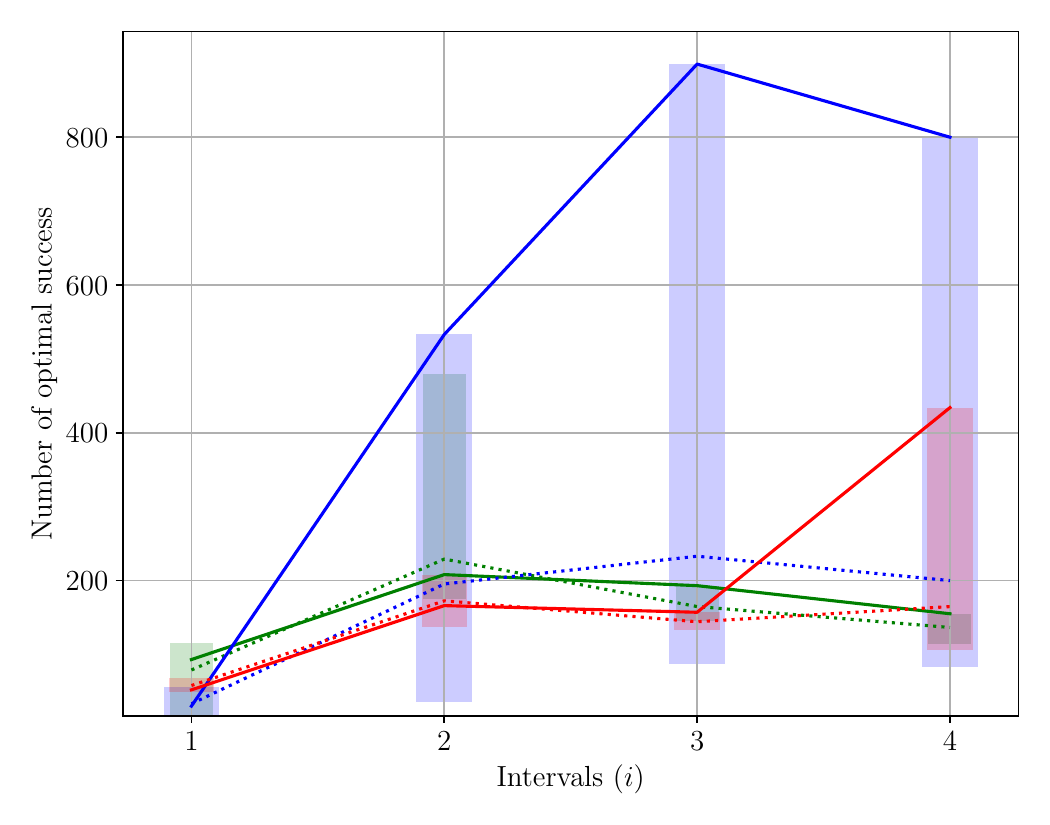} }}%
	\caption{\small In (a) the probability of successful circuits and in (b) the total number of optimal successful circuits in finding a 3-qubit maximally entangled state are noticeably higher with KAN than MLP. A total of 8000 episodes are divided into 4 separate intervals, where each interval contains 2000 episodes. In (a), each point in an interval corresponds to the probability of occurrence of a successful episode (see Eq.~\ref{eq:success_equation}). The results are averaged over 15 random seeds (i.e. initialization) of the networks. The \textit{range} defines the best performance of each interval for both networks. The \textit{range} defines the region between the best and the worst performance in each interval for both networks.}%
	\label{fig:ghz_state_prep_res}%
\end{figure}

In each interval, the probability is averaged over $20$ random seeds, where each seed corresponds to the random initialization of the network. {We observe that the} probability of success with an MLP and KAN is comparable in the first 3 intervals, but in the 4th interval, i.e., in the episode range of $7500-10000$, the probability of success achieved by KAN is higher. With MLP, the best probability of success achievable in the 4th interval is $35.36\%$ whereas with KAN, we can achieve a success probability
$36.31\%$, whereas the number of optimal admissible ansatz achievable by both networks is the same. This indicates that KANs have the potential to generate more diverse solutions to the same problem than MLPs.

Meanwhile, the difference in performance between KAN and MLP becomes significant in the task of constructing the GHZ state. Here we consider a depth 2 {(containing a single hidden layer)} MLP with $3$ neurons ($\textrm{MLP}_{2,3}$) and depth 4 {(containing $3$ hidden layers)} MLP with $30$ neurons ($\textrm{MLP}_{4,30}$) and compare its performance with a depth 2 KAN containing $3$ neurons ($\textrm{KAN}_{2,3}$). We run a total of $8000$ episodes where {in each episode, the RL-agent is allowed to make a quantum circuit of maximum depth $D_\text{max}=12$}. Just like the previous $2$-qubit experiment, we divide the total number of episodes into 4 intervals where each interval contains $2000$ episodes.

In Fig.~\ref{fig:ghz_state_prep_res}(a) we observe that the average probability of success with $\textrm{KAN}_{2,3}$ is higher than both $\textrm{MLP}_{2,3}$ and $\textrm{MLP}_{4,30}$ over $15$ random initialization of the networks. This difference becomes truly significant when we consider the best performance at the 4th interval. \textit{With $\textrm{KAN}_{2,3}$ we can achieve a probability of success in the final interval $48.41\%$ whereas with $\textrm{MLP}_{2,3}$ and $\textrm{MLP}_{4,30}$ we get $38.23\%$ and $38.61\%$ respectively, indicating a $10.18-9.8\%$ performance enhancement with KAN.}

In Fig.~\ref{fig:ghz_state_prep_res}(b), we observe that on average (over $15$ random initialization of the networks), one can achieve a higher number of optimal admissible ansatz using $\textrm{KAN}_{2,3}$ than the MLPs. Meanwhile, when we consider the best performance in the 4th interval \textit{the number of optimal admissible ansatz achievable by KAN is $1.84\times$ to $5.16\times$ higher than $\textrm{MLP}_{2,3}$ and $\textrm{MLP}_{4,30}$ respectively}. {Noticing the significant improvement of RLQAS with KAN in the noiseless scenario, in the following section, we focus on a more realistic setting where quantum gates are subject to noise.}
\begin{center}
\begin{table*}[t]
\centering
\small
\begin{tabular}{ c|c|c|c|c } 
Network & Configuration & Spline and grid & Activation func. & Fidelity \\
\hline
\multirow{1}{4em}{$\text{KAN}_{3,10}$} & $[84, 10,10, 12]$ & B-spline, $k=4,G=5$  & learnable & \textbf{0.7328} \\  
\hline
\multirow{2}{4em}{$\text{MLP}_{4,30}$} 
& $[84, 30,30,30, 12]$ & NA & ReLU & 0.5005\\ 
& & NA & LeakyReLU & 0.7300\\ 
\multirow{3}{4em}{$\text{MLP}_{4,100}$} &  &  & ELU & 0.6830\\ 
& $[84, 100,100,100, 12]$ & NA & ReLU & 0.7300\\ 
&  &  & LeakyReLU & \textbf{0.8500}\\
\hline
\end{tabular}
\caption{\small KAN outperforms most of the configurations of MLPs except when $\text{MLP}_{4,100}$ is operated with LeakyReLU activation function for noisy Bell state preparation with $p_\text{x}=0.3$ and $p_\text{dep}=0.2$. This helps us conclude that to achieve competitive/better performance with an MLP,
it is necessary to fine-tune not just the network’s depth and
width but also the activation function
{In configuration $[84, 10,10, 12]$ the KAN takes $84$ inputs that encode the quantum circuit, calculated using Eq.~\ref{eq:tensor_based_encoding_eqn} for $D_\text{max}=6$, $N=2$ and $N_\text{1q}=5$ and outputs $12$ actions after passing through $2$ hidden layers containing $10$ neurons.}}
\end{table*}
\label{tab:noisy_network_config}
\end{center}

\subsubsection{Noisy simulation} \label{sec:results_noisy}

For the noisy simulation of Bell state preparation, we consider two forms of gate errors. The gate error refers to the imperfection in any quantum operation we perform. For the $1$-qubit gate error, we consider random \texttt{X} noise where with probability $p_\text{x}$ an \texttt{X} gate is applied to the circuit and with $1- p_\text{x}$ it applies an $\mathbb{I}$. Meanwhile, for $2$-qubit gate error, we apply depolarizing noise which replaces the state of any qubit with a random state of probability $p_\text{dep}$. {Under these noisy scenarios, we utilize different configurations of MLP and KAN (summarized in Table~\ref{tab:noisy_network_config}) to construct the Bell state}.

{At the first stage, we consider noise levels $p_\text{x}=0.1$ and $p_\text{dep}=0.01$. Under these circumstances, the $\text{MLP}_{4,30}$ (i.e. an MLP of depth $4$ and $30$ neurons) can achieve a fidelity $99.25\%$ whereas the same fidelity can be achieved with KAN with just depth $2$ and $2$ neurons. Now to elevate the hardness of the problem, in the second experiment, we consider the following noise levels: $p_\text{x}=0.3$ and $p_\text{dep}=0.2$. With KAN configuration presented in Table~\ref{tab:noisy_network_config} we can achieve a fidelity of $73.28\%$ which $\text{MLP}_{4,30}$ with ReLU and LeakyReLU activation functions cannot achieve. Even when the number of neurons is increased tenfold compared to KAN,  $\text{MLP}_{4,100}$ 
(i.e., MLP with depth 4 and 100 neurons) still fails to surpass KAN's fidelity using ELU and ReLU activation functions. However, with LeakyReLU activation, it achieves a higher fidelity of $85\%$.}

\textit{This leads to the conclusion that to achieve competitive/better performance with an MLP, it is necessary to fine-tune not just the network's depth and width but also the activation function. However, with KAN, this process becomes much more straightforward, as the activation functions are learnable}.
\begin{figure}[h!]
    \centering    
    \includegraphics[width=\linewidth]{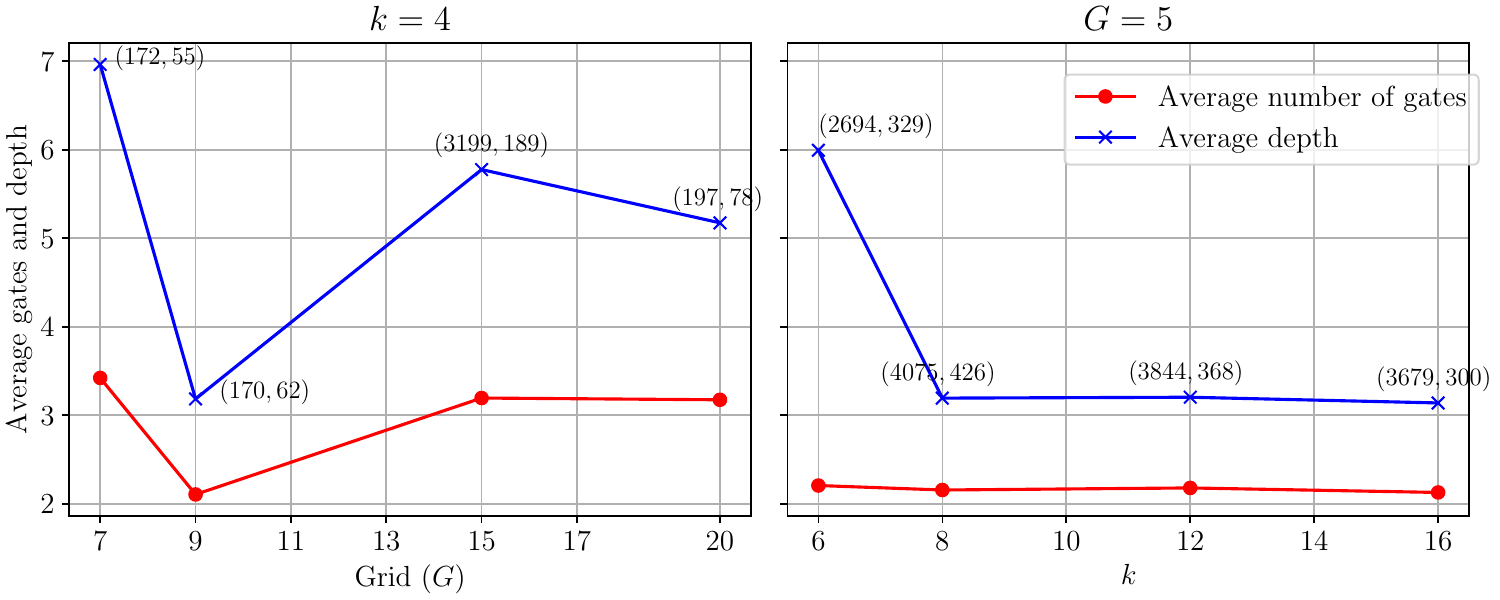}
    \caption{
    \small The number of splines ($k$) has more impact in improving and stabilizing the performance of the KAN than the grid size ($G$). Here we measure the improvement of the KAN while constructing the GHZ state by calculating the average number of gates and depth in each setting. 
    Each point is marked as $(a,b)$ where $a$ is the total number of successful episodes and $b$ total optimal successful episodes.}
\label{fig:succ_k_and_g_variation}
\end{figure}

One can argue that KAN has two additional parameters: splines ($k$) and grid ($G$), and tuning these hyperparameters can heavily affect its performance. In Fig.~\ref{fig:succ_k_and_g_variation}, 
for constructing GHZ state, we show that \textit{to achieve better performance in terms of the depth and number of gates variation in the number of splines (i.e. $k$) is more effective and stable for the network than variation in $G$}. Whence, in ref.~\cite{vaca2024kolmogorov}, the authors show that achieving a better minimization of the loss function grid size within the splines of KANs has a notable impact. 

In the scope of Variational Quantum Algorithms (VQAs) loss function minimization is crucial to achieve a good estimation of a problem, so it is much needed to fine-tune the grid size while optimizing parameterized gates within Parameterized Quantum Circuits (PQCs) of many VQAs. {Following these developments in the upcoming section, we tackle the problem of the construction of effective PQCs that can find the ground state of chemical Hamiltonians with few 2-qubit gates and trainable parameters.}

\subsection{Quantum chemistry} \label{sec:results_chemistry}
{In this section, we utilize a refined version of the recently introduced Curriculum Reinforcement Learning for Quantum Architecture Search (CRLQAS)~\cite{patel2024curriculum} algorithm to automate the search for PQCs for quantum chemistry problems. The refined CRLQAS replaces the MLP structure in the algorithm by KAN, using this construction we evaluate the performance of KAN configurations with MLPs. These neural networks are utilized to find the ground state of 4-qubit $\texttt{H}_2$ and \texttt{LiH} molecules. The molecular structures are discussed in the Tab.~\ref{tab:list_of_mol}.}
\begin{figure*}[t]
    \centering
    \includegraphics[width=0.8\linewidth]{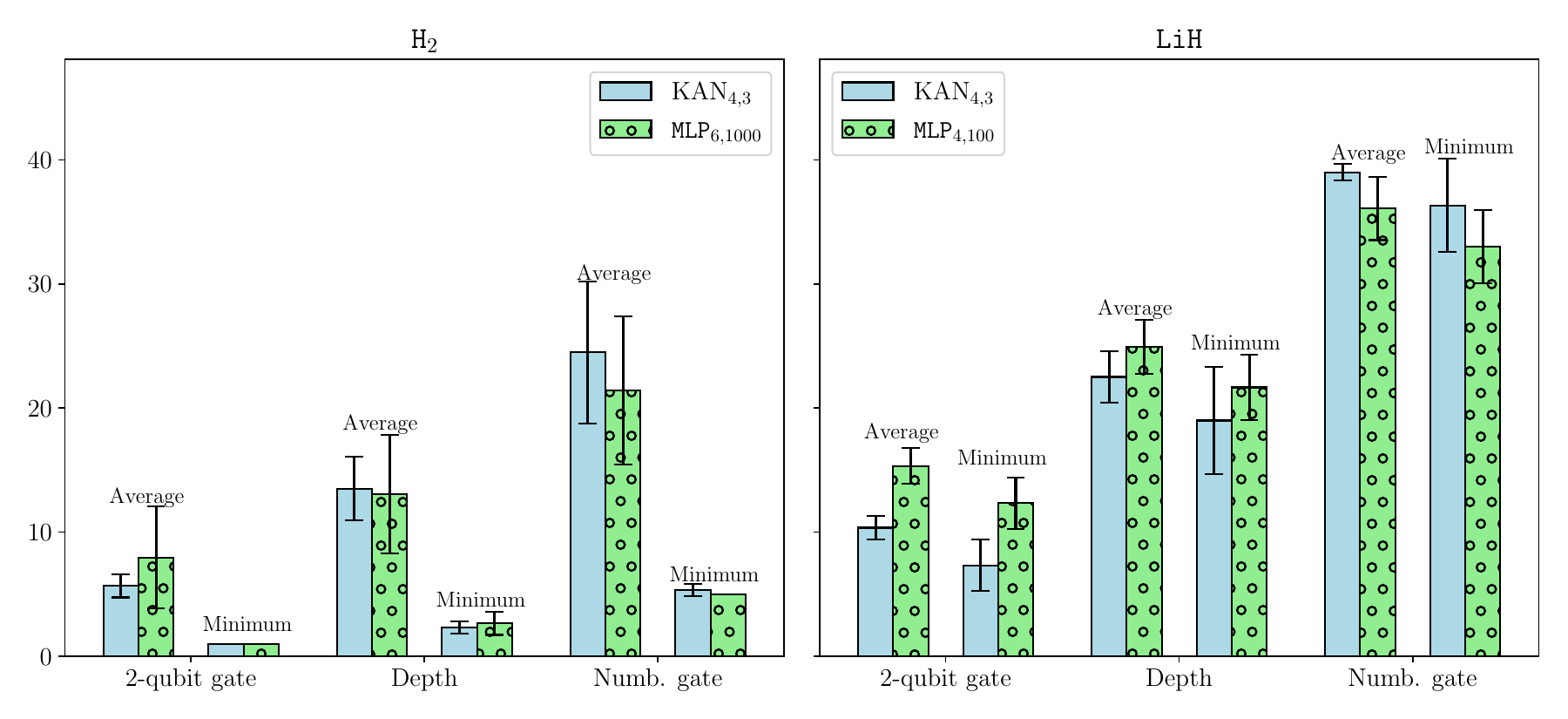}
    \caption{\small \textcolor{black}{KAN outperforms MLP in finding a parameterized quantum circuit that solves the 4-qubit $\texttt{LiH}$ and $\texttt{H}_2$ molecule. We solve a molecule when its energy goes below the chemical accuracy, i.e., $1.6\times10^{-3}$ Hartree. In the plot, the average corresponds to the average performance of the neural networks over three distinct initializations, and the minimum is the best-performing seed among the three. We consider $\text{KAN}_{4,3}$ that corresponds to a KAN of depth 4 and 3 neurons, whereas $\text{MLP}_{6,1000}$ and $\text{MLP}_{4,100}$ defines MLP of depth 6 with 1000 neurons and MLP of depth 4 with 100 neurons respectively. It should be noted that an MLP of configuration $\text{MLP}_{6,1000}$ outperforms KAN in the number of 2-qubit gates, depth, and the total number of parameters, but the number of trainable parameters required for the MLP is $5.15\times10^6$ which is very large compared to $5.52\times10^4$ parameters of KAN.}}
    \label{fig:h2_lih_kan_mlp_compare}
\end{figure*}

{In finding the ground state of $\texttt{H}_2$ molecule, we consider an MLP of depth 6 (equivalently containing 5 hidden layers) with each layer containing 1000 neurons, i.e. $\text{MLP}_{6,1000}$, which is compared with a KAN of depth 4 (equivalently 3 hidden layers) with 3 neurons i.e. $\text{MLP}_{4,3}$. Meanwhile, to find the ground state of \texttt{LiH} molecule we consider $\text{MLP}_{4,100}$ and compare its performance with $\text{KAN}_{4,3}$. In the Appendix~\ref{appndx:quantum_chem_neural_net_config} in Tab.~\ref{tab:noiseless_quant_chem_config} we provide a more detailed discussion of the configuration of the neural networks.}

{
In Fig.~\ref{fig:h2_lih_kan_mlp_compare}, we benchmark the results by calculating the total number of 2-qubit gates, the depth, and the total number of gates in the PQC}. These parameters are labelled as \textit{2-qubit gate}, \textit{Depth}, and \textit{Numb. gate} on the x-axis of Fig.~\ref{fig:h2_lih_kan_mlp_compare}. In both cases, we observe that \textit{$\text{KAN}_{4,3}$ outperforms various MLP configurations in terms of average (as well as minimum) depth and 2-qubit gate count. This indicates that $\text{KAN}_{4,3}$ can find more compact PQCs for solving the $\texttt{H}_2$ and $\texttt{LiH}$ molecules compared to $\text{MLP}_{6,1000}$ and $\text{MLP}_{4,100}$}. However, the total number of gates required to solve these problems is smaller with MLP, suggesting that for a parameterized quantum circuit with fewer 1-qubit gates, it is preferable to choose MLP over KAN in CRLQAS.

\textcolor{black}{In Appendix~\ref{appndx:quantum_chem_results_with_neural_net_config}, we further investigate two different configurations of KAN and MLP to find the ground state of the 4-qubit $\texttt{LiH}$. The results reveal that while larger configurations, such as $\text{MLP}_{6,1000}$, outperform KAN, they require $5\times$ to $100\times$ more learnable parameters. However, when \textit{considering a comparable number of learnable parameters, for instance, $\text{MLP}_{4,500}$ and $\text{KAN}_{4,50}$, KAN achieves the ground state with a shallower PQC and fewer 2-qubit gates}. Although MLP results in a lower error in estimating the ground state, our objective in quantum chemistry is to attain chemical accuracy, defined as $1.6\times10^{-3}$ Hartree\footnote{The chemical accuracy is often defined as being within 1 kcal/mol, which is an experimental value, as 1 Hartree = 627.5095 kcal/mol, therefore 1 kcal/mol is equal to approximately 0.0015934 $\approx$ 0.0016 Hartree}~\cite{chen2020ground}. \textit{Given that the noise in 2-qubit gates is significantly higher than in 1-qubit gates~\cite{wilson2021empirical,johnstun2021understanding}, KAN demonstrates considerable promise for solving quantum chemistry problems in the presence of real quantum hardware noise}.}

\section{Resource details} \label{sec:results_resource}
{Here we quantify the resources such as (1) the total number of learnable parameters and (2) the time of execution of each episode required by KAN and MLP for the simulations presented in the previous section.}

\subsection{Parameter count} 
\begin{center}
\begin{table*}[t]
\centering
\small
\begin{tabular}{ c|c|c|c|c } 
\hline
Problem & Network & Configuration & Spline and grid & Parameter count \\
\hline
\multirow{2}{8em}{Bell state prep. (noiseless)} & KAN & $\text{KAN}_{2,2}$ & B-spline, $k=3,G=5$ &  \textcolor{black}{\textbf{1728}}\\ 
&MLP&$\text{MLP}_{4,30}$ & NA &  \textcolor{black}{4782} \\  
\hline
\multirow{2}{8em}{GHZ state prep. (noiseless)} & KAN & $\text{KAN}_{2,3}$ & B-spline, $k=3,G=5$ & \textcolor{black}{\textbf{5562}}\\ 
&MLP
&  $\text{MLP}_{4,30}$ & NA & \textcolor{black}{$11181$}\\  
\hline
\multirow{2}{8em}{Bell state prep. (noisy)} & KAN &  $\text{KAN}_{3,10}$ & B-spline, $k=4,G=5$ & \textcolor{black}{\textbf{9540}}\\ 
&MLP
&  $\text{MLP}_{4,100}$ & NA & \textcolor{black}{$29912$}\\
\hline
\multirow{2}{8em}{$\texttt{H}_2$ ground state} & KAN &  $\text{KAN}_{4,3}$ & B-spline, $k=4,G=5$ & \textcolor{black}{$\mathbf{5.525\times10^4}$}\\ 
&MLP
& $\text{MLP}_{6,1000}$ & NA & \textcolor{black}{$5.150\times10^6$}\\
\hline
\multirow{2}{8em}{$\texttt{LiH}$ ground state} & KAN & $\text{KAN}_{4,3}$ & B-spline, $k=4,G=5$ & \textcolor{black}{$\mathbf{5.525\times10^4}$}\\ 
&MLP
& $\text{MLP}_{4,100}$ & NA & \textcolor{black}{$1.348\times10^5$}\\
\hline
\end{tabular}
\caption{\textcolor{black}{KAN requires $2$-$3\times$ less learnable parameters in the task of quantum state preparation. Meanwhile, in finding the ground state of $\texttt{LiH}$ and $\texttt{H}_2$ KAN takes $2.4$-$93\times$ fewer parameters respectively and still solves the problems with smaller parameterized quantum circuits than MLP. To simplify, we denote the networks by their depth and number of neurons. For instance, a KAN with configuration $[288,3,21]$ is abbreviated as $\text{KAN}_{2,3}$, where 2 represents the network depth (excluding the input layer) and 3 denotes the number of neurons in the hidden layer. {Meanwhile, $288$ is the size of the input calculated using Eq.~\ref{eq:tensor_based_encoding_eqn} for $D_\text{max}=12$, $N=3$, and $N_\text{1q}=5$, and $21$ is the number of actions}. The same shorthand notation is applied to MLP networks.}}
\label{tab:num_learnable_params}
\end{table*}
\end{center}
{With depth $L$ and width $N$, an MLP} only requires $O(N^2L)$ parameters whereas a KAN requires $O(N^2L(G+k))$ parameters. In Table~\ref{tab:num_learnable_params} for the quantum state preparation and in Tab.~\ref{tab:quant_chem_diff_config_result} for quantum chemistry problems, we calculate the total number of learnable parameters required for KANs and MLPs in noisy and noiseless scenarios. \textit{We observe that in all cases KAN requires a lesser number of parameters than MLP}.

\subsection{Time of executing each episode}
\textcolor{black}{Although KAN requires fewer learnable parameters than MLPs, the average time of executing each episode for KAN is $120\times$ higher. \textit{A recently released version of KAN, named MultKAN~\cite{multkan}, executes an episode approximately $3.46\times$ slower than MLPs as shown in Tab.~\ref{tab:kan_time}}}.
\begin{center}
\begin{table*}[t]
\centering
\small
\begin{tabular}{ c|c|c|c } 
\hline
Problem & Network & Configuration & Avg. time per episode  \\
\hline
\multirow{2}{8em}{GHZ state prep.} & KAN & $\text{KAN}_{2,3}$ & {0.2049} \\
&MLP& $\text{MLP}_{4,30}$ & \textbf{0.0592}\\  
\hline
\multirow{2}{8em}{$\texttt{H}_2$ ground state} & {KAN} & $\text{KAN}_{4,3}$  & {2.2103}\\
&MLP& $\text{MLP}_{6,1000}$ & \textbf{0.7552} \\
\hline
\multirow{2}{8em}{$\texttt{LiH}$ ground state} & {KAN} & $\text{KAN}_{4,3}$  & {3.3737}\\
&MLP& $\text{MLP}_{4,100}$ & \textbf{1.2102} \\
\hline
\end{tabular}
\caption{\small \textcolor{black}{A new version of KAN~\cite{multkan} is approximately $3.46\times$ slower than MLPs, which is far better than the older version, which was $120\times$ slower for quantum state preparation. Meanwhile, KAN is $2.92\times$ and $2.787\times$ slower than MLP in finding the ground state of 4-qubit $\texttt{H}_2$ and $\texttt{LiH}$ molecule respectively.}}
\label{tab:kan_time}
\end{table*}
\end{center}

\section{Discussion and future work} \label{sec:discussion}
{In this work we analyze the practicality of the question: \textit{Can Kolmogorov-Arnold Network (KAN) replace the Multi-Layer Perceptron (MLP) structure in Quantum Architecture Search (QAS) to achieve comparable or better performance while mitigating the issues of parameter growth and activation function selection? We demonstrate that the answer is affirmative by tackling the quantum state preparation task and finding the ground state of chemical Hamiltonians}. To this end, we propose KAN for QAS, namely KANQAS, which replaces the MLP in the reinforcement learning-assisted quantum architecture search of the double deep Q-network with the KAN.}

Our experiments reveal that \textit{KANQAS can increase the probability of success of the RL-agent in finding optimal quantum circuits compared to MLPs in constructing multi-qubit maximally entangled states with non-parameterized gates}. Moreover, \textit{when noise is present in the quantum device, achieving similar or better outcomes with an MLP necessitates a deeper and wider network compared to KAN, as well as a careful selection of the appropriate activation function. Since finding the optimal activation function for deep learning remains an ongoing challenge~\cite{hayou2018selection, ramachandran2017searching}, KAN has an advantage, as its activation functions are inherently learnable}.

\textcolor{black}{\textit{In addressing quantum chemistry problems, KAN demonstrates the ability to provide a more compact parameterized quantum circuit, with less number of 2 qubit gate and depth for finding the ground state of the chemical Hamiltonian using VQE, with a significantly reduced number of learnable parameters}. It is important to note that for larger configurations, such as $\text{MLP}_{6,1000}$, MLP outperforms KAN configurations but requires $5$-$100\times$ more learnable parameters. When \textit{considering a comparable number of learnable parameters, for example, $\text{MLP}_{4,500}$ and $\text{KAN}_{4,50}$, KAN achieves the ground state with a parameterized quantum circuit of smaller depth and fewer 2-qubit gates. Although MLP results in a lower error in estimating the ground state, our objective in quantum chemistry is to attain chemical accuracy, defined as $1.6\times10^{-3}$ Hartree. Given that the noise in 2-qubit gates is significantly higher than in 1-qubit gates~\cite{wilson2021empirical}, KAN demonstrates considerable promise for solving quantum chemistry problems on real quantum hardware}.}

Although the number of learnable parameters in KAN is on average $2$-$100\times$ lesser than in MLPs, one of the biggest disadvantages of KAN is that it requires $2$-$4\times$ more execution time per episode as compared to MLPs. However, due to their effectiveness and efficiency in finding solutions in noiseless and noisy quantum devices, KAN thus appears to be a reasonable alternative to traditional MLPs in solving quantum architecture search problems. In the following, we discuss the \textit{future research directions} as a follow-up to our research.
\begin{itemize}
    \item \textbf{KAN for VQAs:} A primary direction for future research is addressing the quantum architecture search problem within variational quantum algorithms, \textcolor{black}{in this paper we just show the simulation of 4-qubit $\texttt{H}_2$ and $\texttt{LiH}$ molecules, this study should be expanded to bigger molecules such as $\texttt{H}_2\texttt{O}$ 8-qubits. These KAN-assisted algorithms could have significant applications in quantum chemistry and combinatorial optimization problems}.
    \item \textbf{Optimizing computational time of KAN:} Another important goal is to explore the use of specialized accelerators, such as tensor processing units or digital signal processors, to reduce the computation time of KAN.
    \item \textbf{Interpretability of KAN:} Focusing on the interpretability of KAN, future research should investigate its applicability to similar but scalable problems to enhance understanding. This can, for example, include the investigation of subclasses of families of the activation function after training KAN towards concept discovery~\cite{sarkar2024yaqq}. \textcolor{black}{\textit{We invite readers to explore our Git repository~\cite{kanqas_code} for inspiration and further motivation in this direction}}.
\end{itemize}
\section{List of Abbreviations}
\textbf{KANQAS:} Kolmogorov-Arnold Network for Quantum Architecture Search; \textbf{KAN:} Kolmogorov-Arnold Network; \textbf{MLP:} Multi-Layer Perceptron; \textbf{QAS:} Quantum Architecture Search; \textbf{CRL:} Curriculum Reinforcement Learning; \textbf{RL:} Reinforcement Learning; \textbf{VQA:} Variational Quantum Algorithms; \textbf{NN:} Neural Network; \textbf{NISQ:} Noisy Intermediate-Scale Quantum; \textbf{DDQN:} Double Deep Q-Network; \textbf{CRLQAS:} Curriculum Reinforcement Learning for Quantum Architecture Search; \textbf{GHZ:} Green-berger–Horne–Zeilinger; \textbf{VQE:} Variational Quantum Eigensolver; \textbf{PQC:} Parameterized Quantum Circuit.

\section{Declarations}
\subsection{Ethical Approval and Consent to participate}
Ethical approval for this study was obtained from the authors, and all the authors provided informed consent to participate.
\subsection{Consent for publication}
The authors grant permission for the publication of this research.
\subsection{Availability of supporting data}
The supporting data is available from the corresponding author upon reasonable request. The code to produce the data is available at \href{https://github.com/Aqasch/KANQAS_code}{https://github.com/Aqasch/KANQAS\_code}.
\subsection{Competing interests}
The authors declare no competing interests.
\subsection{Authors' contributions}
A.K. conceptualized the study, supervised the project, wrote the code, analyzed the results, and prepared the manuscript.
A.S. analyzed the results and contributed to manuscript preparation.
A. Sadhu analyzed and prepared the results, and participated in manuscript preparation.
\subsection{Funding}
A.K. acknowledges funding from the Research Council of Finland through the Finnish Quantum Flagship project 358878 (UH).

\textcolor{black}{%
\section{Acknowledgement}
AK would like to thank Ziming Liu for fruitful discussions about KAN.
AK wish to thank the Finnish Computing Competence Infrastructure (FCCI) for supporting this project with computational and data storage resources.}

\section{Appendix}
\begin{center}
\begin{table*}[t]
\centering
\small
\begin{tabular}{ c|c|c|c|c } 
\hline
Network & Problem & Configuration & Spline and grid & Activation func. \\
\hline
\multirow{2}{4em}{KAN} & Bell state prep. &  $\textrm{KAN}_{2,2}$ & B-spline, $k=3,G=5$ & learnable \\ 
&GHZ state prep. &  $\textrm{KAN}_{2,3}$ & B-spline, $k=4,G=5$ & learnable  \\  
\hline
\multirow{3}{4em}{MLP} & Bell state prep. &  $\textrm{MLP}_{4,30}$ & & \\ 
&GHZ state prep.
&  $\textrm{KAN}_{2,3}$ & NA & LeakyReLU \\  
& GHZ state prep. &  $\textrm{KAN}_{4,30}$ & &\\
\hline
\end{tabular}
\caption{\small Configuration for noiseless GHZ and Bell state preparation. In the case of MLP, the activation function is fixed whereas it is learned during the network training for KAN. For the sake of simplicity, we represent the networks by their depth and number of neurons {for example, for a KAN of configuration $[288,3,21]$ we use the shorthand notation $\text{KAN}_{2,3}$ as the KAN is of depth 2 (i.e. 1 hidden layer) and contains 3 neurons in the hidden layer. The KAN takes $288$ size input which is the size of the tensor that encodes the quantum state calculated using Eq.~\ref{eq:tensor_based_encoding_eqn} for $D_\text{max}=12$, $N=3$, and $N_\text{1q}=5$, and returns $21$ actions.}}
\label{tab:noiseless_state_prep_network_config}
\end{table*}
\end{center}
{
\subsection{Tensor-based binary encoding of parametric quantum circuit}\label{appndix:tensor_based_encoding}
\begin{figure}[H]
	\includegraphics[width=\linewidth]{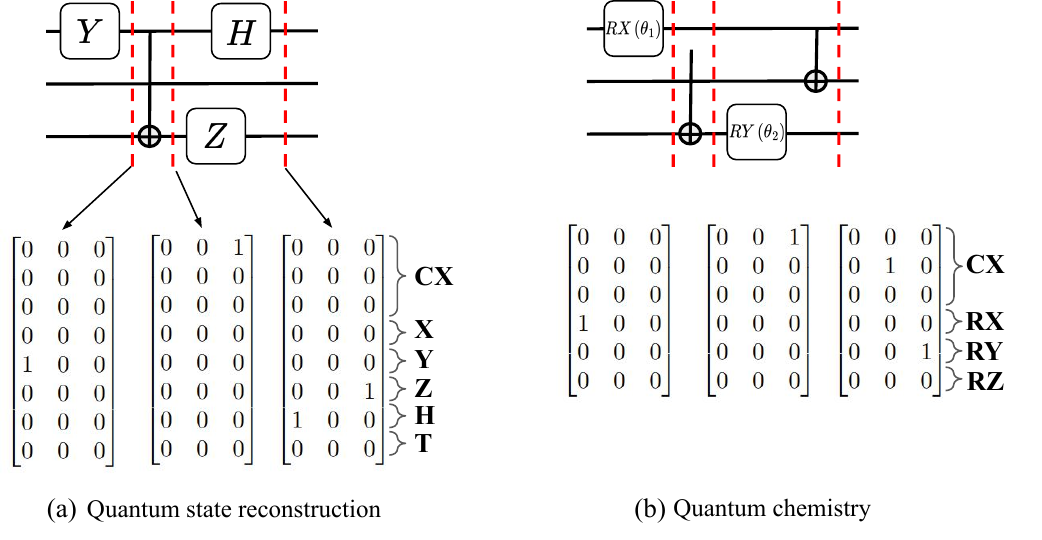}
	\caption{{(a) For quantum state reconstruction the number of 1-qubit gates is one 5 so we encode the quantum circuit into a tensor of dimension $\left[D_{\text{max}} \times \left((N+5)\times N\right)\right]$ (b) whereas, for the quantum chemistry problem, the PQC is encoded into a tensor of dimension $\left[D_{\text{max}} \times \left((N+3)\times N\right)\right]$ due to the presence of 3 1-qubit gates.}}
	\label{fig:tensor_based_encoding}
\end{figure}
We use a binary encoding scheme~\cite{patel2024curriculum,kundu2024enhancing} that captures the structure of the Parametric Quantum Circuit (PQC), specifically the order of gates, to provide the agent with a full circuit description. To maintain a constant input size, the tensor is prepared for the deepest possible quantum circuit.}

{To construct the tensor, we first set the hyperparameter $D_{\text{max}}$, which limits the maximum number of allowed gates (actions) in all episodes. A \textit{moment} in a PQC refers to all gates that can be executed simultaneously, defining the circuit's depth.}

{We represent PQCs as 3D tensors where, at each episode, we initialize an empty circuit of depth $D_{\text{max}}$, defined by a $\left[D_{\text{max}} \times \left((N+N_\text{1q})\times N\right)\right]$ tensor of zeros, where $N$ is the number of qubits, and $N_\text{1q}$ is the number of 1-qubit gates. Each matrix in the tensor has $N$ rows for control and target qubit positions in \texttt{CNOT} gates, followed by either 3 (for quantum chemistry problems) or 5 (for quantum state reconstruction problems) rows indicating the positions of 1-qubit gates as represented in Fig.~\ref{fig:tensor_based_encoding}.}

\subsection{Double Deep Q-Network (DDQN)}\label{appndx:ddqn}
Deep Reinforcement Learning (RL) techniques utilize Neural Networks (NNs) to adapt the agent's policy to optimize the return:
\begin{eqnarray}
G_t = \sum_{k=0}^{\infty} \gamma^k r_{t+k+1},
\end{eqnarray}
where $\gamma \in [0,1)$ is the discount factor. For each state-action pair $(s,a)$, an action value is assigned, quantifying the expected return from state $s$ at step $t$ when taking action $a$ under policy $\pi$:
\begin{eqnarray}
q_\pi (s,a) = \mathbbm{E}_\pi [G_t | s_t = s, a_t = a].
\end{eqnarray}

The goal is to determine the optimal policy that maximizes the expected return, which can be derived from the optimal action-value function $q_\ast$, defined by the Bellman optimality equation:
\begin{eqnarray}
q_\ast (s,a) = \mathbbm{E} \bigg[ r_{t+1} + \max_{a'} q_\ast (s_{t+1},a') | s_t = s, a_t = a \bigg].
\end{eqnarray}

Instead of solving the Bellman optimality equation directly, value-based RL aims to learn the optimal action-value function from data samples. Q-learning is a prominent value-based RL algorithm, where each state-action pair $(s,a)$ is assigned a Q-value $Q(s,a)$, which is updated to approximate $q_\ast$. Starting from randomly initialized values, the Q-values are updated according to the rule:
\begin{eqnarray}
&&Q(s_t,a_t) \gets Q(s_t,a_t) \nonumber \\
&& \qquad \qquad + \alpha \bigg( r_{t+1} + \gamma \max_{a'} Q(s_{t+1},a') - Q(s_t,a_t) \bigg),  \nonumber \\
\end{eqnarray}
where $\alpha$ is the learning rate, $r_{t+1}$ is the reward at time $t+1$, and $s_{t+1}$ is the state encountered after taking action $a_t$ in state $s_t$. This update rule is proven to converge to the optimal Q-values in the tabular case if all $(s,a)$ pairs are visited infinitely often~\cite{melo2001convergence}. To ensure sufficient exploration in a Q-learning setting, an $\epsilon$-greedy policy is used. Formally, the policy is defined as:
\begin{align}
\pi(a|s)\coloneqq \begin{cases}
1-\epsilon_t & \text{if $a = \max_{a'} Q(s,a')$},\\
\epsilon_t & \text{otherwise}.
\end{cases}
\end{align}
\begin{center}
\begin{table*}[t]
\centering
\small
\begin{tabular}{ c|c|c|c|c } 
\hline
Network & Molecule & Network configuration & Spline and grid & Activation func. \\
\hline
\multirow{4}{4em}{KAN} & $\texttt{H}_2$ & $\text{KAN}_{4,3}$& \\
& & $\text{KAN}_{2,3}$ & B-spline, $k=10,G=5$ & learnable  \\
& $\texttt{LiH}$ & $\text{KAN}_{4,30}$ &  \\
&  & $\text{KAN}_{4,50}$ &  \\
\hline
\multirow{4}{4em}{MLP} & $\texttt{H}_2$ & $\text{MLP}_{6,1000}$ & & \\ 
& & $\text{MLP}_{4,100}$ & & \\
& $\texttt{LiH}$ & $\text{MLP}_{4,500}$ & NA & LeakyReLU \\
& & $\text{MLP}_{6,1000}$ & & \\
\hline
\end{tabular}
\caption{\small \textcolor{black}{The configuration of KAN and MLP utilized in tackling quantum chemistry problems specifically for finding the ground state of 4-qubit $\texttt{H}_2$ and $\texttt{LiH}$ molecule. A detailed configuration of these molecules is in Tab.~\ref{tab:list_of_mol}. As discussed in the caption of Tab.~\ref{tab:noiseless_state_prep_network_config}, we represent the networks by their depth and number of neurons for example, for an MLP of configuration $[1121,500,500,500,24]$, we use the shorthand notation $\text{MLP}_{4,500}$ as the MLP of depth 4 (excluding the input layer) and of 500 neurons in the hidden layer.}}
\label{tab:noiseless_quant_chem_config}
\end{table*}
\end{center}
\begin{center}
\begin{table*}[t]
\centering
\small
\begin{tabular}{ c|c|c|c|c|c|c } 
\hline
Molecule & \makecell{Network \\ configuration} & \makecell{Min. 2-qubit \\ gate} & Min. depth & \makecell{Min. 1-qubit\\ gate} & error & \makecell{Parameter\\ count} \\
\hline
\multirow{4}{4em}{4-qubit $\texttt{LiH}$}
&  $\text{KAN}_{4,30}$ & 11 & 18 & 15 & $1.1\times10^{-4}$ & $\mathbf{5.8\times 10^{5}}$ \\
&  $\text{KAN}_{4,50}$ & 8 & 17 & 24 & $2.2\times10^{-4}$ & $9.9\times 10^{5}$ \\
&  $\text{MLP}_{4,500}$ & 13 & 18 & 13 & $1.1\times10^{-4}$ & $1.0\times 10^{6}$ \\
&  $\text{MLP}_{6,1000}$ & \textbf{6} & \textbf{11} & \textbf{13} &$\mathbf{3.6\times10^{-5}}$ & $5.1\times 10^{6}$ \\
\hline
\end{tabular}
\caption{\small 
\textcolor{black}{We observe that although $\text{MLP}_{6,1000}$ gives the best results in all benchmarking variables, it requires $10\times$ more learnable parameters than $\text{KAN}_{4,30}$, making it significantly more expensive in terms of training time. For a fair comparison, $\text{KAN}_{4,50}$ and $\text{MLP}_{4,500}$ show a tradeoff between the number of parameterized 1-qubit gates and 2-qubit gates, with $\text{MLP}_{4,500}$ having 1.078 times more parameters than $\text{KAN}_{4,50}$. Despite having fewer parameters, KAN provides a competitive approximation of the ground state with a shorter parameterized quantum circuit, fewer 2-qubit gates, and smaller depth than MLP.}}
\label{tab:quant_chem_diff_config_result}
\end{table*}
\end{center}

The $\epsilon$-greedy policy introduces randomness to the actions during training, but after training, a deterministic policy is used.

We employ NN ad function approximations to extend Q-learning to large state and action spaces. NN training
typically requires independently and identically distributed data. This problem is circumvented by experience replay. This method divides experiences into single-episode updates and creates batches which are randomly sampled from memory. For stabilizing training, two NNs are used: a policy network, which is continuously updated, and a target network, which is an earlier copy of the policy network. The current value is estimated by the policy network, while the target network provides a more stable target value $Y$ given by :
\begin{eqnarray}
Y_{\text{DQN}} = r_{t+1} + \gamma \max_{a'} Q_{\text{target}} (s_{t+1},a').
\end{eqnarray}

In the DDQN algorithm, we sample the action for the target value from the policy network to reduce the overestimation bias present in standard DQN. The corresponding target is defined as:
\begin{eqnarray}
Y_{\text{DDQN}} = r_{t+1} + \gamma Q_{\text{target}} \bigg( s_{t+1}, \arg \max_{a'} Q_{\text{policy}} (s_{t+1},a') \bigg).
\end{eqnarray}
This target value is approximated via a loss function. In this work, we consider the loss function as the smooth L1-norm

\subsection{Configuration of neural networks}\label{appndx:quantum_chem_neural_net_config}
The detailed configurations of the KANs and the MLPs utilized in the main text to tackle quantum state preparation and quantum chemistry problems are provided in Tab.~\ref{tab:noiseless_state_prep_network_config} and Tab.~\ref{tab:noiseless_quant_chem_config}

\subsection{\textcolor{black}{Quantum chemistry with different configurations of KAN and MLP}}~\label{appndx:quantum_chem_results_with_neural_net_config}

\textcolor{black}{In this section, we further explore the different configurations of KAN and MLP in finding the ground state of the 4-qubit $\texttt{LiH}$ molecule. We use the following variables to benchmark the performance of the different neural networks: the minimum number of 1-qubit gates (Min. 1-qubit gate), the minimum number of 2-qubit gates (Min 2-qubit gate), the error in estimating the ground state energy (error), and the number of learnable parameters in the network (Parameter count). The results are presented in detail in the Tab.~\ref{tab:quant_chem_diff_config_result}}.

\textcolor{black}{From the Tab.~\ref{tab:quant_chem_diff_config_result} we observe although the $\text{MLP}_{6,1000}$ provides us with the best result in terms of all the benchmarking variables, the number of learnable parameters it requires to achieve this outcome is $10\times$ higher than $\text{KAN}_{4,30}$, which is far more expensive in terms of training time. Meanwhile, for the sake of fair comparison, if we observe the performance of $\text{KAN}_{4,50}$ and $\text{MLP}_{4,500}$ where the number of parameters in MLP is $1.078\times$ more than KAN, we notice a tradeoff between the number of parameterized 1-qubit gates and the number of 2-qubit gates. The KAN provides us with a competitive approximation of the ground state with a shorter parameterized quantum circuit, containing less 2-qubit gate and smaller depth than MLP despite having a smaller number of parameters. Considering that the noise in 2-qubit gates in real quantum hardware is many scales higher than 1-qubit gates, KAN with a comparable number of parameters, shows a large potential in quantum architecture search in quantum hardware.}

\clearpage
\newpage

\clearpage
\newpage

\bibliographystyle{ieee}
\bibliography{ref.bib}

\end{document}